\documentclass[]{aa} 
\usepackage{graphicx}
\usepackage{txfonts}
\begin{document}
\title{A VLT/FLAMES survey for massive binaries in Westerlund 1. III. The  WC9d binary W239 and implications for massive  stellar 
evolution.\thanks{Based 
on observations made at the European Southern Observatory, Paranal,
Chile under programs ESO 81.D-0324 \& 383.D-0633 and OPD/LNA Observatory (Br).}
} 
\author{J.~S.~Clark\inst{1}
\and B.~W.~Ritchie\inst{1}
\and I.~Negueruela\inst{2}
\and P.~A.~Crowther\inst{3}
\and A.~Damineli\inst{4}
\and F.~J.~Jablonski\inst{5}
\and N.~Langer\inst{6}
}
\institute{
$^1$Department of Physics and Astronomy, The Open 
University, Walton Hall, Milton Keynes, MK7 6AA, United Kingdom\\
$^2$Departamento de F\'{i}sica, Ingenier\'{\i}a de Sistemas y Teor\'{i}a de la Se\~{n}al, Universidad de Alicante, Apdo. 99,
E03080 Alicante, Spain\\ 
$^3$Department of Physics \& Astronomy, Hicks Building, University of Sheffield, Hounsfield Road, Sheffield, S3 7RH, United Kingdom\\ 
$^4$ Instituto de Astronomia, Geof\'{i}sica e Ci\^{e}ncias Atmosf\'{e}ricas, Universidade de S\~{a}o Paulo, Rua do Mat\~{a}o 1226, Cidade 
Universit\'{a}ria, 05508-090, S\~{a}o Paulo, SP, Brazil\\
$^5$Instituto Nacional de Pesquisas Espaciais/MCT, Avenida dos Astronautas 1758, S\~{a}o Jos\'{e} dos Campos, SP, 12227-010, Brazil
$^6$Argelander-Institut f\"{u}r Astronomie der Universit\"{a}t Bonn, Auf dem H\"{u}gel 71, 53121 Bonn, Germany
}

   \abstract{There is growing evidence that  a treatment of binarity amongst OB stars is essential for a full theory of stellar evolution. However the binary properties of massive stars - frequency, mass ratio \& orbital separation - are still poorly constrained.} 
{In order to address this shortcoming  we have  undertaken a  multiepoch spectroscopic study of the stellar population  of the young massive cluster Westerlund~1. In this paper
we present an investigation into the nature of the  dusty Wolf-Rayet star and  candidate binary  W239.} 
{To accomplish this we have utilised our spectroscopic data in conjunction with multi-year optical and near-IR photometric observations in order to search for binary 
signatures. Comparison of these data to synthetic  non-LTE model atmosphere spectra were used to derive the fundamental properties of the WC9 primary.}
{We found W239 to have an orbital period of only $\sim$5.05~days, making it one of the most compact   WC binaries yet identified. Analysis of the long term 
near-IR lightcurve reveals 
a significant flare between 2004-6. We interpret this as evidence for a third massive stellar component in the system in a long 
period ($>$6~yr), eccentric orbit, with dust production occuring at periastron leading to the flare. The presence of
a near-IR excess characteristic of hot ($\sim$1300~K) dust at every epoch  is consistent with the expectation that the 
subset of persistent dust forming WC stars are  short ($<$1~yr) period binaries, although confirmation will require further observations.
Non-LTE model atmosphere analysis of the spectrum reveals the physical properties of the WC9 component to be fully consistent with other Galactic examples. }
{The simultaneous presence of both  short period Wolf-Rayet binaries and cool hypergiants within Wd~1 provides compelling evidence 
for a bifurcation
in the post-Main Sequence evolution of massive stars due to binarity. Short period O+OB binaries will evolve directly to the Wolf-Rayet phase, either  due to an episode of binary mediated mass 
loss - likely via case A mass transfer or a contact configuration -  or via chemically homogenous evolution. 
Conversely, long period binaries and single stars will instead undergo a red loop across the HR diagram via a cool hypergiant phase.
 Future analysis of the  full spectroscopic dataset for Wd~1 will  
constrain the proportion of massive stars experiencing each pathway; hence quantifying the importance
 of binarity in massive stellar evolution up to and beyond supernova and the resultant production of relativistic 
remnants and X-ray binaries.}


\titlerunning{The WC9d binary W239}
\maketitle

\section{Introduction}

Recently, there has been increasing recognition of the importance of
binarity in the evolution of massive stars. The binary fraction of
such objects is expected to provide an important observational test of
current theories of star formation (e.g.  Bonnell \& Bate
\cite{bonnell}; Zinnecker \& Yorke \cite{zin}) and the early (rapid)
dynamical evolution of young massive stellar clusters. As both binary
components evolve it is thought that substantial quantities of mass
may be lost in close binaries as the primary expands to (over)fill its
Roche Lobe (e.g. Petrovic et al. \cite{petrovica}). Such binary driven
mass loss is also predicted to influence the nature of both subsequent
supernova (e.g. Paczynski et al.  \cite{pacz}; Podsiadlowski et
al. \cite{pod}) and resultant relativistic remnant (Wellstein \&
Langer \cite{wellstein}; Fryer et al.  \cite{fryer}). Moreover, recent
evolutionary simulations suggest that in very compact binaries tidal
interaction may drive homogeneous chemical evolution permitting the
formation of very massive stellar mass black holes (de Mink et
al. \cite{demink}) while it is also suspected that a proportion of
Gamma Ray Bursts may result from binary evolution channels (Cantiello
et al. \cite{cantiello}).

Powerful observational support for the importance of binary mediated
mass loss and homogeneous chemical evolution has been provided by the
dynamical determination of the masses of both components of galactic
(4U1700-37 \& GX301-2 -- Clark et al. \cite{clark02a}; Kaper et
al. \cite{kaper}) and extragalactic (IC10 X-1 \& NGC300 X-1 --
Crowther et al. \cite{ic10}; Silverman \& Filippenko et
al. \cite{silverman}) high mass X-ray binaries. However, the relative
importance of these pathways depends on both the overall binary
percentage amongst massive stars as well as the distribution of
orbital separations and mass ratios (e.g. Kobulnicky \& Fryer
\cite{kobulnicky}). Consequently significant effort has been applied
to the determination of these properties via multiepoch (multiplexed)
spectroscopic observations of massive stellar aggregates (e.g. Cyg
OB2, NGC6231 \& 30 Dor -- Kiminki et al. \cite{kiminki}; Sana et
al. \cite{sana}; Bosch et al. \cite{bosch}) whereby homogeneous
stellar populations may be efficiently sampled.  These studies reveal
a uniformly high binary fraction for OB stars ($\geq$40\%), which,
when corrected for the observational bias against long period systems
results in a fraction consistent with unity for at least one region
(Cyg OB2; Kobulnicky \& Fryer \cite{kobulnicky}).

Another compelling target is the young (4-5~Myr) massive
($\sim$10$^5$~M$_{\odot}$) galactic cluster Westerlund 1 (Wd~1; Clark
et al. \cite{clark05}), which contains a significant ($>>$100)
population of massive (M$_{initial}>$30~M$_{\odot}$; Ritchie et al.
\cite{ben10}) stars. Multiwavelength observations of the cluster
revealed a number of indirect binary markers amongst the OB
supergiants and Wolf-Rayets (WR), with a binary fraction for the
latter inferred to be at least 70\% (and potentially consistent with
unity; Crowther et al. \cite{pac06}; Clark et al. \cite{clark08};
Dougherty et al. \cite{dougherty}). Motivated by these findings we
have undertaken a systematic radial velocity (RV) spectroscopic survey
to identify and characterise the binary population of Wd~1; a detailed
description of the project methodology, reduction technique and
preliminary results may be found in Ritchie et al. (\cite{ben09}).

In this paper we present an analysis of the properties of the WC9d
star W239 (=WR F, WR77n), for which binarity was suggested by the
presence of a hard X-ray spectrum and a pronounced near-mid IR excess
due to warm circumstellar dust (Skinner et al. \cite{skwd1}; Crowther
et al. \cite{pac06}). Including W239, 7 of the 8 WCL (= WC8-9) stars
within Wd1 demonstrate one or both of these observational properties.
Such phenomena are typically attributed to the wind collision zone in
a massive binary system where material is first shock heated to high
temperatures, with the concommitant high densities subsequently
permitting downstream dust condensation upon cooling (e.g. Prilutskii
\& Usov \cite{PU}; Allen et al. \cite{allen}; Usov et al. \cite{usov};
Williams \& van der Hucht \cite{williams92}).

We present a brief description of the spectroscopic and photometric
datasets employed in this work in Sect. 2, as well as a full RV
solution - to the best of our knowledge only the fifth for a dusty WC
binary after WR 113, 137 \& 140 (Massey \& Niemela \cite{mass113};
Williams et al. \cite{williams90}; Lef\`{e}vre et al. \cite{lef}) and the LMC system 
HD 36402 (=Br31; Moffat et al. \cite{moffatLMC}, Williams  \cite{perry11}).
The results of a tailored non-LTE atmospheric analysis of the WR
component of the binary is also included in Sect. 2, while we discuss
the implications of our findings in Sect. 3, before summarising them
in Sect. 4.

\section{Data Reduction \& Analysis}

\subsection{Spectroscopy}

Eleven epochs of spectra were obtained between 2008-9 with the Fibre
Large Array Multi Element Spectrograph (FLAMES; Pasquini et
al. \cite{pas}) mounted on VLT UT2, with the GIRAFFE spectrograph
operated in MEDUSA mode (HR21 setup), resulting in a resolving power
of $\sim$16200 between 8484-9001~$\AA$. Full details of data reduction
and analysis may be found in Ritchie et al. (\cite{ben09}).

While Clark et al. (\cite{clark10}) found that W239 was not subject to
secular variability between 2001-9, Ritchie et al. (\cite{ben09})
report the presence of RV variations in the C\,{\sc iii} 8500 \&
8664~$\AA$ lines in the 2008 spectra (e.g. Fig. 1). As a result RV
measurements of all 11 epochs of observations were carried out using
the Nelder-Mead simplex method to fit Lorentzian profiles to the
strong C\,{\sc iii} lines, providing a robust RV determination that is
unaffected by small-scale line-profile variations due to wind effects.
The results of this process are summarised in Table 1, with sample
fits presented in Fig. 1.  

 An obvious explanation for the pronounced
RV shifts is that they are due to reflex motion induced by an unseen
companion to the WC9d primary. A  period search of the RV data strongly
favours   a candidate period of ~5.05days. An alias of this period at 3.04
days may not be excluded from the current data but results in larger (O-C)
residuals and appears further disfavoured because wind-driven mass loss in
the WC phase  acts to  widen the orbit, implying   that W239 would have had
to have passed  through a particularly   compact configuration prior to its
current state;  we return to this  point in Sect.~3.3.1. Consequently, we
adopt a period of 5.05 days for the remainder of this paper and  emphasise
that the  analysis  explicitly excludes longer  orbital periods ($>6$~days)
and that any residual uncertainty in the period {\em does not} materially
affect our conclusions.

 We show the resultant
curve derived from a Levenberg-Marquardt fit to the RV data in Fig. 2,
with an (orbital) period,
P$_{orb}$=5.053$\pm$0.002~days\footnote{Errors are internal to the
  fit.}, and semi-amplitude K$_{WC}$=39.7$\pm$4.8
kms$^{-1}$ (Table 2). Eccentricity was fixed at zero, due to the short
  period and expectation that binary interaction would have
  circularized the orbit. We note that deviations from the RV fit are
  greater in the 2009 data ($\sim$10--15kms$^{-1}$), suggestive of
  either a small systematic error in the period determination or wind
  variability superimposed on the orbital motion.  The systemic
velocity derived from the C\,{\sc iii} 8664$\AA$ line is -60.5$\pm$3.5
kms$^{-1}$, with the C\,{\sc iii} 8500$\AA$ line offset redwards as
noted by Ritchie et al. (\cite{ben09}). While slightly higher than the
mean derived from 10 supergiants by Mengel \& Tacconi-Garman
(\cite{mengel}) it is within the range found for other cluster members
from our full FLAMES dataset (Ritchie et al. \cite{ben09},
\cite{ben10b}).

It might, however, be supposed that such line profile variability was
the result of rotating global wind structures.  Indeed, periodic
spectral variability attributed to such a cause has previously been
reported in the optical lines of the apparently single WRs WR1 \& 6
(Morel et al. \cite{morel}; Chen\'{e} \& St-Louis
\cite{chene}). However, we consider this unlikely for several
reasons. Firstly, the RV line profile variability in W239 encompasses
the whole profile.  This contrasts to the behaviour of WR 1 \& 6,
where superposition of additional transient emission components on a
stationary underlying profile leads to shifts in the line centre of
gravity or `skewness'. This is also the case for the apparently single
WC8 star WR K within Wd1, for which we find line profile variability
in the line core but no global RV shifts (Ritchie et al., in prep.).
Secondly, the 5.05~day period present in W239 is coherent over
$\sim$83 cycles; in comparison, the quasi-period of WR 1, attributed
to wind asymmetries, is only stable for $\leq$4 cycles (Chen\'{e} \&
St.-Louis \cite{chene}).  Finally, this wind variability is
accompanied by photometric modulation with an identical period in both
WR 1 \& 6, but Bonanos (\cite{bonanos}) demonstrate that such
behaviour is absent in W239.

Moreover, phase resolved near-IR spectral analysis of 5 WR binaries
reveals that variability due to wind-wind interaction driven
asymmetries - rather than binary reflex motion - is in any event
phased on their known orbital periods (Stevens \& Howarth
\cite{stevens}).  We are thus confident that the 5.05~day period
identified in the RV curve of W239 reflects an underlying binary
orbital period of the system.

Motivated by this finding we re-examined the FLAMES spectra of W239 to
search for the signature of a companion(s), noting that the shorter
wavelength data were of too low S/N and resolution to permit such an
analysis and the longer wavelength data are dominated by a near-IR
excess due to hot dust (Sect. 2.2 \& 2.3). The FLAMES spectra cover
Pa11-16, which are expected to be in absorption in late O/early B
stars. Unfortunately Pa11 falls on the red wing of a C\,{\sc iv}
emission line from the WC9 primary, while Pa13 and Pa16 fall at the
same wavelengths as strong C\,{\sc iii} lines. Given the decrease in
the strength of the Paschen series to higher transitions, Pa12
(8750{\AA}) appears to be the best remaining binary diagnostic.
Unfortunately, while this transition is clear of any WR emission lines
it does overlap the interstellar C$_2$ Phillips (2-0) R/Q bands, which
are visible in our spectra of W239 and other cluster members,
complicating analysis (cf. Ritchie et al. \cite{ben10}). This region
of the spectra is presented in Fig. 3, and we note that W239 appears
to share the weak, broad Pa12 profile found for this transition (and
the wider Paschen series) in the O+O spectroscopic binary W30a (Clark
et al. \cite{clark08}; Negueruela et al. \cite{w30}); we therefore
regard this as {\em tentative} evidence for a hot massive
companion(s). Unfortunately, the weakness and breadth of the line,
combined with the low S/N and presence of interstellar features
precludes a search for RV variability, although pre-empting Sect. 2.3
\& 3.1 one might expect any reflex motion in a putative early type
companion to be of low amplitude, such that it would be difficult to
identify at the resolution of these data.
 
\begin{table}
\caption{Journal of observations.}
\label{tab:observations}
\begin{center}
\begin{tabular}{ll|cc}
\hline
\hline
 MJD$^{a}$   & Phase (Orbits)$^{b}$ & RV(C\,{\sc iii} 8664) & O-C\\
& & (kms$^{-1}$)$^c$ \\
\hline
 54646.1846 & 0.000           & $-47.4$ & $-1.0$ \\
 54665.0356 & 0.731~(3.73)    & $-31.1$ & $-0.5$ \\
 54671.1343 & 0.938~(4.94)    & $-29.2$ & $5.8$ \\
 54692.0423 & 0.076~(9.08)    & $-62.1$ & $1.4$ \\
 54713.0107 & 0.226~(13.23)   & $-87.9$ & $4.6$ \\
 54724.0818 & 0.417~(15.42)   & $-89.8$ & $2.0$ \\
 54728.0554 & 0.203~(16.20)   & $-91.0$ & $-1.8$ \\
 54734.0613 & 0.392~(17.39)   & $-85.5$ & $9.3$ \\
 54965.1768 & 0.132~(63.13)   & $-92.6$ & $-16.6$ \\
 54969.3198 & 0.952~(63.95)   & $-22.2$ & $15.0$ \\
 55063.0575 & 0.503~(82.50)   & $-87.1$ & $-10.5$ \\
\hline
\end{tabular}
\end{center}
$^{a}$Modified Julian day at the midpoint of two 600s integrations
(2$\times$500s, 04/09/2008; 1$\times$600s+1$\times$700s, 19/09/2008),
$^{b}$Phase taking $T_0$=54646.1846, elapsed orbits for an orbital
period of 5.053d, $^c$RV listed for the C\,{\sc iii} 8664$\AA$ line, 
with fitting errors $\le4$km~$s^{-1}$ at all epochs.\\
\end{table}

\begin{table}
\caption{Orbital parameters}
\label{tab:orbital}
\begin{center}
\begin{tabular}{lc}
\hline\hline
Parameter & Value \\
\hline
$P$ (days)                    & $5.053\pm0.002$ \\
$K_{WC}$ (km~s$^{-1}$)          & $39.7\pm4.8$    \\
$\gamma$ (km~s$^{-1}$)         & $-60.5\pm3.5$ \\
$i$                           & $< 30^\circ$ \\
$e$                           & 0 (fixed) \\
$a$~sin~$i$ (R$_\odot$)        & $3.97\pm0.48$\\
\hline
\end{tabular}
\end{center}
\end{table}

\subsection{Photometry}

\begin{table}
\caption{Summary of near-IR photometry}
\begin{center}
\begin{tabular}{lcc}
\hline
\hline
Date     & H & nbK \\
\hline
19/05/99 & 7.61 & - \\
28/06/04 & 8.21 & - \\
29/06/04 & -    & 6.78 \\ 
19/06/04 & 8.18 & - \\
24/09/04 & 7.97 & 7.28 \\
03/06/05 & 7.69 & 5.78 \\
27/07/05 & 7.67 & - \\
28/07/05 & 7.64 & - \\
29/07/05 & 7.63 & - \\
31/07/05 & 7.64 & 5.08 \\
12/08/05 & 7.73 & - \\
16/07/06 & 7.56 & - \\
04/08/06 & 7.59 & - \\
05/08/06 & 7.52 & - \\
06/08/06 & 7.59 & - \\
08/08/06 & 7.53 & - \\
09/08/06 & 7.60 & - \\
10/08/06 & 7.55 & - \\
31/05/07 & -    & 6.34 \\
05/09/09 & 7.67 & - \\
06/09/09 & 7.67 & - \\
18/06/09 & 7.63 & - \\
30/04/10 & 7.77 & 6.29 \\
19/07/10 & 7.68 & - \\
20/07/10 & 7.65 & - \\
21/07/10 & 7.68 & - \\
22/07/10 & 7.68 & - \\
\hline
\end{tabular}
\end{center}
The errors on H band data of $\pm$0.05mag 
and on nbK band of $\pm$0.17mag, with the exception of the 
data from 24/09/04 where they are $\pm$0.05 in both bands (Crowther et al. 
\cite{pac06}).
\end{table} 

Wd~1 has been the subject of a long term
broad band  I, H \& K and narrow band K\footnote{henceforth nbK, centred at 2.14$\mu$m, FWHM=0.023$\mu$m.} continuum 
 observational campaign since 2004, utilising the 0.6-m and 1.6-m telescopes at Pico dos
 Dias Observatory (LNA/Brazil). These were equiped with  an Hawaii 1024$\times$1024 pixel camera, which gave a field-of-view of 
8$\times$8~arcmin and 4$\times$4~arcmin respectively. 
Standard image processing techniques were adopted (bias subtraction, 
flatfielding and sky subtraction) prior to data extraction via aperture photometry under  IRAF.
Finally, absolute I band magnitudes were obtained via  calibration to the data presented by Piatti 
et al. (\cite{piatti}), while the near-IR photometry was calibrated via comparison to 2MASS data.

Unfortunately, W239 was saturated in the broad K band, leaving just 5 epochs of observations  in the 
narrowband 2.14$\mu$m filter. However 25 individual H band observations between 2004-10 were viable; both datasets
  are plotted in Fig. 4 and summarised in Table 3, along with  photometry  from Crowther et al. 
(\cite{pac06}). These are further supplemented with the I band data from Piatti 
et al. (\cite{piatti}) and Clark et al. (\cite{clark05}) and 10 epochs of new observations. 
Unfortunately,  archival 2MASS data were associated with large uncertainties and so were not employed.

 These data  provide no evidence for secular variability  in the I band between 1995-2010. However, dramatic 
near-IR variability is present, with a significant brightening  observed between 2004-6 ($\Delta$H$\sim$0.6~mag, $\Delta$nbK$\sim$2.2~mag).
This was also accompanied by the system becoming redder, with (H-nbK)$\sim$0.69$\pm$0.07 on  2004 May 1 and 2.6$\pm$0.2 on 2005 July 31. 
 Subsequently, W239 appears to have faded in the nbK band, albeit to a higher quiescent 
 flux than found during the initial phase of the outburst - with the inevitable corollaries that the data are 
both  sparse and are associated with comparatively large uncertainties. In contrast the H band luminosity was 
observed to increase by a smaller amount, but appears to have remainded at the peak luminosity for the 
following 4 year period. This was also accompanied by a modest reduction to (H-nbK)$\sim$1.3$\pm$0.2 by 2010.
Unfortunately, lack of contemporaneous data did not permit us to determine if the outbursts peaked at the same time in both 
bands.

Similar photometric behaviour - a transient outburst characterised by an increase in  IR continuum  excess with   wavelength - 
is observed in  other  WC binary systems where dust production rates are modulated on the orbital period and peak during periaston passage 
(e.g. Williams et al. \cite{williams94}, \cite{williams01}, van der Hucht et al. \cite{vdh01}). Consequently, we 
infer the presence of an additional long period tertiary companion to W239, with the lack of a second outburst suggesting 
a period of at least $\sim$6~yr for this component. However, the  broadband near-IR colours   
{\em at all times} are clearly indicative of the continued presence of hot  dust (e.g. Crowther et al. \cite{pac06}); in this respect
W239 bears a close resemblance to WR48a, which also shows both persistent and episodic, enhanced dust emission.
Williams et al. (\cite{williams03}) also interpret this  as evidence for a triple system, with a 
short period binary (P$_{\rm orb}<$1~yr) component responsible for  continous dust formation and an 
interaction with a tertiary companion  in  a  wide, eccentric orbit (P$_{\rm orb}>$30~yr; P. Williams, priv. comm. 2010) 
leading to the outburst in 1980. We return to this in Sect. 3.

However, in this regard we highlight that despite the sparse sampling, the decay in flux from the outburst peak in the nBK band lightcurve appeared to occur more 
rapidly than that of the the H band. In other periodic/episodic dust producing systems the opposite is observed, with more extended decays at
 progressively longer wavelenths due to  the cooling of the dust as it is carried from the system by the stellar winds. Unfortuntely, given the 
limitations of our current dataset - in particular the lack of longer wavelength data - we cannot presently  explain this apparent difference in behaviour. 

Finally,  optical observations by  Bonanos (\cite{bonanos}) between 2006 June 15-July 25 revealed 
no evidence for eclipses or elipsoidal modulation in the lightcurve. We repeated the experiment with a higher 
cadence - $\sim$1500 individual I band observations between 2008 July 8-17 - but again found no evidence for 
modulation on the orbital period although, as with Bonanos, intra-night variability of   $\Delta$I$\leq$~0.08mag. 
over short ($\sim$hr) timescales was observed.

\begin{figure}
\resizebox{\hsize}{!}{\includegraphics[angle=0]{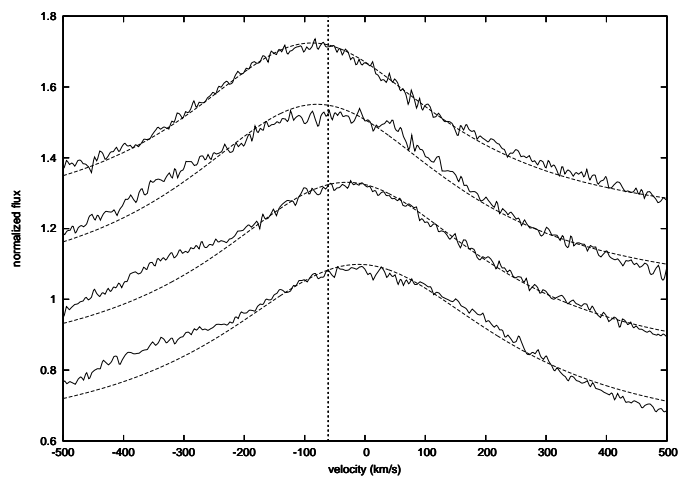}}
\caption{Top panel: sample C\,{\sc iii} 8664{\AA} emission line profiles for W239 clearly indicating the 
binary induced radial velocity  shift relative to the systemic velocity (dashed line).  
Lorentzian profile fit to emission line as utilised for the radial velocity determination overplotted. Note the presence of an additional
C\,{\sc iii} emission feature in the blue wing at 8653.2~{\AA}; this was found to have no effect on the RV determination.
}
\end{figure}

\begin{figure}
\resizebox{\hsize}{!}{\includegraphics[angle=0]{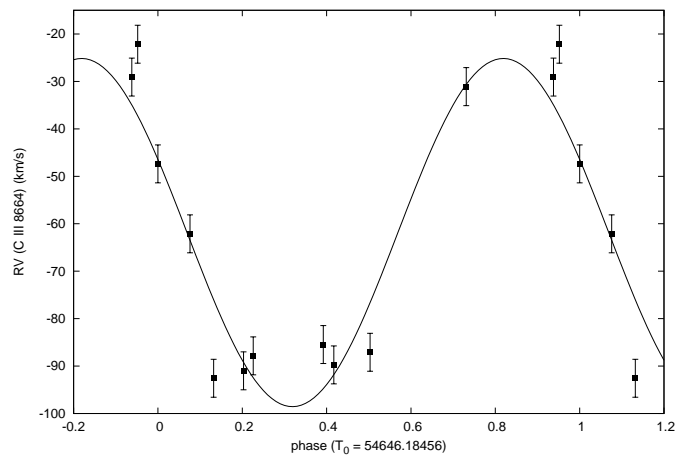}}
\caption{Radial velocity curve for W239, phased on the orbital period. Velocity determined from the C\,{\sc iii} 8664~{\AA} line.}
\end{figure}

\begin{figure}
\resizebox{\hsize}{!}{\includegraphics[angle=0]{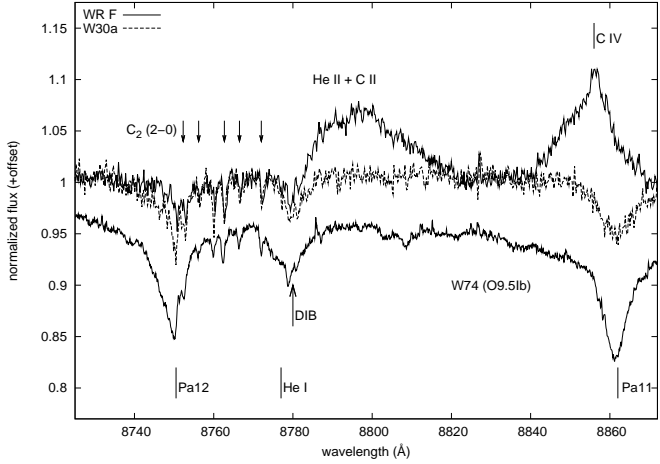}}
\caption{Comparison of the 8700-8900~{\AA} spectra of W239 (solid line) and the O+O SB2 binary W30a (dotted line). 
W239 appears to demonstrate a similar weak, broad Pa12 profile at $\sim$8500~{\AA} as W30a. The weakness of this feature is apparent upon consideration of the single 
O9.5 Ib cluster member W74.}
\end{figure}

\begin{figure}
\resizebox{\hsize}{!}{\includegraphics[angle=270]{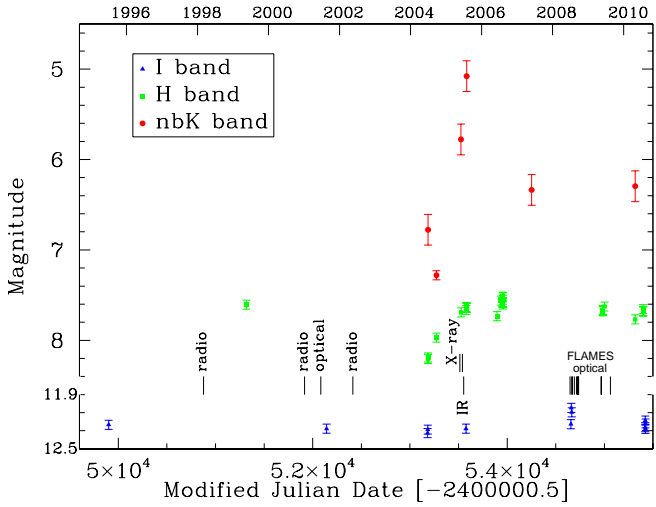}}
\caption{Optical-near IR lightcurve for W239. The locations of optical and near-IR spectroscopic, radio 
and X-ray observations discussed in the text are also indicated.}
\end{figure}

\begin{figure}
\includegraphics[width=6cm,angle=-90,bb=55 30 550 780]{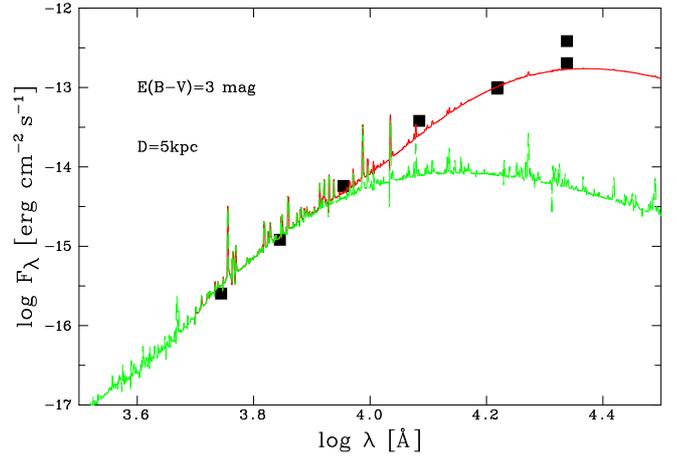}
\caption{Comparison between the broadband photometry of W239 and the reddened WC+O spectral energy distribution 
excluding (green dashed line) and including (red solid line) the simple 1300~K graybody dust model.
Optical photometry from Clark et al. (\cite{clark05}; 2001 August 21) and the near-IR from 2005
June 3 \& July 31 (this work).}
\end{figure}

\begin{figure}
\includegraphics[width=8.5cm,bb=30 0 545 780]{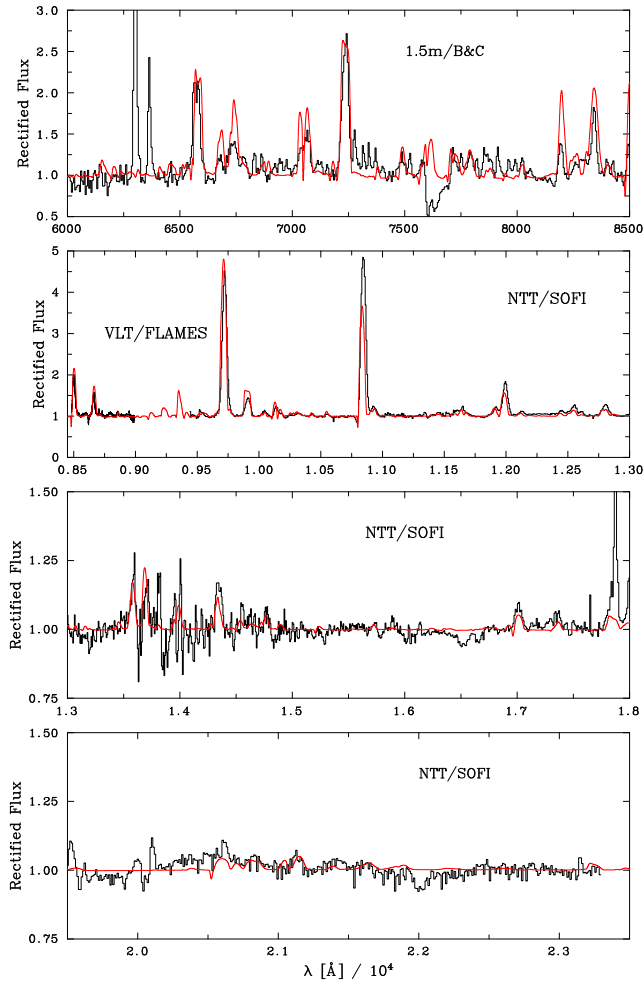}
\caption{Comparison of the rectified synthetic spectrum of W239 (red line) to observations
(black line), acounting for the WC9 line dilution from the unseen companion and warm dust. Spectra were obtained in 2001 
June (1.5m/B\&C), 2005 June (NTT/SofI) and 2008 July (VLT/FLAMES). Note the two emission features
shortwards of 6500{\AA} are the result of incomplete sky subtraction; telluric contamination is also manifest in the observed spectrum 
 beween $\sim$7200-7400~{\AA}, 1.35-1.50~$\mu$m and 1.77-1.8~$\mu$m.}
\end{figure}

\subsection{The physical properties of W239}

In order to investigate the nature of the WC9 primary of W239 we have employed the non-LTE model atmosphere code 
CMFGEN  (Hillier \& Miller \cite{hillier98}, \cite{hillier99}). To accomplish this, in addition to the data 
described in Sect. 2.1 we also  employed the spectra presented by 
Clark et al. (\cite{clark02}) and Crowther et al. (\cite{pac06}) and we refer the reader to those works 
for a description of  the observational and reduction procedures employed. While these data
span a significant period of time, no secular spectral evolution is apparent (Sect. 2.1) and there is an excellent 
correspondence between regions of the spectrum that have been multiply sampled.
The methodology adopted   
follows that of the analysis  of the WC9 star WR103 by Crowther et al. (\cite{wc9}). Given the significant reddening to W239 some of the UV/optical transitions 
utilised by Crowther  et al. were unavailable to us; thus our primary  
diagnostics were  C\,{\sc ii}$\lambda\lambda$ 6678, 7230, 9900~{\AA}, C\,{\sc iii }$\lambda\lambda$ 
  6740, 8500, 9700, 21100~{\AA} and C\,{\sc iv}$\lambda$ 14300~{\AA} which, when combined with local continuum levels  
enabled an estimation of temperature, wind properties and stellar luminosity.

One complication was the significant dilution of the spectrum at long ($>1$~$\mu$m)  wavelengths by emission from hot 
dust, as well as a potential  further contribution from a massive companion(s). The former was 
accounted for by the inclusion of a simple greybody model for the dust emission, assuming a single temperature of  
1300~K. To represent  the latter we utilised a Kurucz LTE model with T$\sim$35~kK - corresponding to an O7 V star - and log$g \sim$4. 
We  adopted an  equal light ratio at 8500~$\AA$ for the WR and OB companion(s). The relative contribution of the O7~V star was determined
from the analysis of Martins \& Plez (\cite{plez}) who found M$_V \sim$-4.7  and we adopted (V-I)$_0 \sim$-0.47 (appropriate for an  O star; Cox et al. \cite{cox})
 which yielded M$_I \sim$-4.23, while our WC9 model has M$_V \sim$-4.3 and
(V-I)$_0 \sim$-0.05 leading to M$_I \sim$-4.25.

The results of this model are presented in Table 4.  Encouragingly, they bear close resemblance to the parameters of other 
WC9 stars analysed in an identical manner (Crowther et al. \cite{pac02}, \cite{wc9}). The resulting best fit 
synthetic  spectrum (Figs. 5 \& 6) provides an acceptable   match to the spectrum from the  optical to near-IR   ($\sim$0.6-2.2~$\mu$m) and 
photometry for V through K for a distance, d = 5 kpc and reddening,   E(B-V) = 3. The resultant composite synthetic 
spectrum demonstrates no {\em pronounced} spectral features from the putative companion; consistent with  the current observational data.
Unfortunately, W239 is saturated and/or blended in both {\em Midcourse Source 
Experiment} (MSX)   and {\em Spitzer} Galactic plane surveys; hence more sophisticated modeling of the IR  excess, 
although of considerable interest, appears unwarranted at this time. 
At longer wavelengths, the 3.6~cm radio flux predicted from the model is a factor $\sim$5 smaller than observed ($\sim$0.3~mJy; Dougherty et al. \cite{dougherty}), suggesting
the possibility of an additional contribution resulting from the wind collision.  Finally, we note that our inability to {\em unambiguously}
assign contributions to the continuum from the primary and companion(s) precludes us determining an 
accurate current mass for the WC9 star via the H-free mass-luminosity calibration of Schaerer \& Maeder (\cite{ml}), although we 
consider 10~M$_{\odot}$ to be a conservative {\em upper} limit.

\begin{table}
\begin{center}
\caption[]{Summary of physical properties of the WC9 primary of W239}
\begin{tabular}{lc}
\hline
\hline
Property & Value \\
\hline
$T_*$ (kK) & 52 \\
log($L/L_{\odot}$) & 4.95 \\
$R_*/R_{\odot}$ & 3.8 \\
log($\dot{M}$/$\sqrt{f}$) (M$_{\odot}$ yr$^{-1}$) & -4.5 \\ 
{\em f}       & 0.1 \\
$v_{\infty}$ (kms$^{-1}$) & 950 \\
$X_C$     & 0.25 \\
\hline
\end{tabular}
\end{center}
\end{table}

\section{Discussion}

\subsection{The nature of the W239 system}

The combination of cyclic RV variability, an IR excess, hard X-ray emission when single WC stars are not known to emit 
at such energies (e.g. Oskinova et al. \cite{os03}) and the tentative spectroscopic detection of the luminous companion(s) 
provides compelling evidence for the multiplicity of W239. Indeed, the mismatch between
 the 5.05~day RV periodicity and long term photometric behaviour is suggestive of a triple system. But what can be deduced about 
the putative companions from a synthesis of these observational datasets?

Unfortunately, the nature of the binary  companion in the 5.05~day  orbit is somewhat uncertain (in the absence of an RV curve from 
the putative Pa12 absorption line we may not assign it to either a binary or tertiary component with certainty).  Nevertheless, assuming a 
mass for the WC9 component of $\sim$8-10~M$_{\odot}$ (Sect. 2.3), the combination of the observed RV semi-amplitude, K$_{WC}$, 
lack of eclipses and stellar radius derived from modeling imply a lower limit to the
 mass for the unseen companion of $\sim$2~M$_{\odot}$ (for an inclination of $\sim$45$^{\rm o}$). However based on evolutionary considerations - specifically the 
requirements placed on the system mass ratio for 
merger to be avoided during 
interaction (Sect. 3.3) - we suspect the actual companion mass is significantly larger, with the 
 consequent constraint that the binary is observed at low inclination.

As an {\em illustrative} example, an hypothetical  10~M$_{\odot}$ (WC9) + 20 -- 30~M$_{\odot}$ (B0 -- O7 V) binary 
observed at $\sim$30 -- 10$^{\rm o}$ would replicate the observed K$_{WC}$=32~kms$^{-1}$.
 A  putative O7 V  companion would be consistent with  our {\em assumption} that the WC9 primary and 
companion contribute 
equally at 8500~{\AA}, while a fainter B0 V star  would also satisfy this condition while also accomodating   the presence of a tertiary companion of comparable spectral type.
The  resultant orbital separation of only $\leq$0.2~AU   would then be an order of magnitude smaller than that of the archetypal dust producer 
WR140 (P$_{\rm orb}\sim$7.94~yr) during periastron. We emphasise that neither scenario represents an attempt to provide a fully 
self consistent model for the W239 system; a goal which must await future, stricter observational constraints. 

Would  such a putative binary  be consistent with the current X-ray data for W239? In Table 5 we summarise the properties of known WC binary systems displaying
phenomena associated with wind collision zones (X-ray and/or dust production). Unfortunately, no counterpart to an X-ray bright WC9 binary has yet been 
classified, although the WC8+O7.5 III binary WR11\footnote{The 
physical properties of the WC8 star in WR11 - $T_*=$57.1~kK, log($L/L_{\odot}$)=5.0, log($\
dot{M}$/$\sqrt{f}$)=-4.53, 
$v_{\infty}$=1550~kms$^{-1}$  (De Marco et al. \cite{dem}) - are similar to those of W239, save for  a faster wind. In contrast, 
earlier WC stars are significant  hotter, more luminous and support winds with higher velocity and mass loss rates 
(Crowther et al. \cite{pac02}).} suggests that  companions of  comparatively late spectral type may yield detectable 
X-ray emission in conjunction with WCL stars,  although we note that this system in {\em not} a dust producer.

With L$_X \sim 2\times10^{32}$~ergs$^{-1}$, W239 is an order of magnitude fainter than  WR11 (Clark et al. \cite{clark08}), but is also a much more
compact system. For the binary  configuration described above, the orbital separation is only an order of magnitude larger than the stellar radii of the 
WC9 and O7~V stars (Martins et al. \cite{martins}); hence one would  expect that neither wind will have reached terminal velocity at the position of the shock, 
likely reducing the resultant X-ray luminosity in comparison to WR11. Moreover, one might also expect radiative braking of the WR wind by the O star 
(cf. V444 Cyg; Owocki \& Gayley \cite{gayley}), further suppressing X-ray emission. 

 The X-ray observations of W239 were made in two blocks, one of 18~ks ($\sim$4\% of orbit) on 2005 May 22  and  42~ks ($\sim$10\% of orbit) on 
2005 June 20, a separation of 5.8 cycles (Clark et al. \cite{clark08}). No variability in flux was found within or between 
either observation. While X-ray variability as a function of orbital phase has been observed in WR11 (e.g. Schild et al. 
\cite{wr11}) that system is observed at higher inclination, such that significant orbital variation in the wind properties and hence line of sight extinction
are present, which would be absent for the putative low inclination configuration proposed here. Consequently, one can at least qualitatively understand the X-ray 
properties of W239, although quantitative confirmation would require a tailored hydrodynamical simulation, which currently appears unwarranted given the lack of 
observational constraints.

Finally, we briefly turn to the nature of the putative P$_{orb} >$6~yr tertiary component suggested by the IR lightcurve. Unfortunately, the lack of contemporaneous
 spectroscopy over this period precludes any firm conclusions, save that a $\leq$O7 V or $\sim$O9 III star are suggested by the twin constraints imposed by the 
age of Wd1 and the requirement of the star to support a substantive wind to permit  dust production, noting that  we might also expect such a star 
to contribute to the observed X-ray emission.

\subsection{Dust production in close binaries}

Our current IR observations are consistent with the hypothesis that hot dust is always present within the W239 system.
Monnier et al. (\cite{monnier}) suggested that the subset of the persistently dusty WC stars were 
short orbital period ($<1$~yr) binaries, where dust formation occurs in  long lived Archimedean spirals that result
from high density  material produced at the wind collision zone being carried outwards by a uniformly  expanding wind 
until temperatures are low enough to permit condensation (Tuthill et al., \cite{tuthill99}).
With a  $\sim$5.05~day orbital period -  comparable to that of HD 36402 in the LMC (Moffat et al. \cite{moffatLMC}, 
Williams \cite{perry11}) and a factor of $\sim$6 smaller than  WR113\footnote{Two other dust WC9d
stars - WR69 \& 103 - have also been suggested to be short period binaries on 
the basis of periodic photometric variability. For the former star Marchenko et al. (\cite{hipp}) report a 
 2.29~day periodicity from Hipparcos data but are unable to determine its origin (binarity or NRP), although 
earlier observations by Balona et al. (\cite{balona}) failed to  identify this. Conversely, both
 the latter authors and Moffat et al. 
(\cite{moffat}) report a $\sim$1.75~day photometric period for WR 103 which was not found in the Hipparcos dataset. Given 
the potentially transient nature of the periodicity in both stars and the lack of a confirmatory 
RV curve we regard the adoption of these periods as orbital in origin as premature.}, the next most compact dust forming WC 
binary in the Galaxy - W239 potentially presents an ideal testbed for such a prediction.

 Hydrodynamical simulations  of the wind collision zones by Pittard (\cite{pittard})  indicate that coherent large scale 
Archimedean spirals may form in massive short period systems, although their  properties are highly dependent on the binary 
separation. Specifically, the large scale wind structure of the  most compact binary considered by Pittard 
(\cite{pittard}; P$_{orb} \sim$3~days) quickly fragmented, leading to a highly clumpy, inhomogeneous outflow. Could 
dust formation  occur in such an environment? Unfortunately, {\em ab initio} simulations of such a 
process are not yet available and are certainly premature for W239. 
However,   radiative transfer models of the wider (P$_{orb}\sim$241.5~day) binary WR104 of Harries et al. 
(\cite{harries})  suggest that self shadowing from the dense inner regions of wind spiral strongly influences the temperature
structure and hence location of dust condensation; it is unclear  whether such a process would also operate 
in the fragmented wind spiral anticipated for W239, which in turn might be expected to affect 
dust production if the density at which the wind  cooled to below the dust sublimation temperature 
was too low to permit dust grain growth.

In this regard we note with interest that a number of WC+O star CWBs with comparable periods are {\em not} 
dust producers\footnote{WR 9 (WC7+O5-8, P$_{\rm orb} \sim$14.3~day; Niemela \cite{niemela}),  WR 42 (WC7+O7 V, P$_{\rm orb} \sim$7.9~day;
Hill et al. \cite{hill}),   WR79 (WC5+O6, P$_{\rm orb} \sim$8.9~day; Hill et al. \cite{hill}) and the LMC system HD 36521 (WC4+O6 V-III), P$_{\rm orb}
\sim$1.92~day; Moffat et al. \cite{moffatLMC})} although these are all of earlier spectra type (WC4-7).
One might speculate that the production of dust in W239 is a result of  the softer radiation field of the WC9  primary 
when compared to these hotter stars,  increasing the likelihood of  survival of dust grains in this system, although the recent discovery of dust 
asociated with the LMC system HD 36402 (WC4(+O?)+O8 I; Williams \cite{perry11}) potentially presents a counterarguement to this hypothesis.

\begin{table*}
\begin{center}
\caption[]{Summary of dusty WC binary properties (top panel) and remaining X-ray detected binaries (lower panel).}
\begin{tabular}{llllll}
\hline
\hline
WR & Spectral Type & P$_{\rm orb}$ & L$_X$ &Period determinations & Reference \\ 
   &               &           & (10$^{33}$ erg s$^{-1}$) &   & \\
\hline
 & & & & & \\
{\bf 77n/W239} & WC9vd+?     & 5.053$\pm$0.002~day &0.4 & RV & This work \\
     &             &               & &    & Clark et al. (\cite{clark08}) \\
& & & & & \\
19 & WC4pd + O9.6 & 10.1~yr & - & photometric & Veen et al. (\cite{veen}) \\
& & & & & \\
48a &WC8ed + ?    &  $\sim$1~yr + $>$30~yr?  &  100  & photometric (provisional) & Williams et 
al. (\cite{williams03}) \\
            &      &                          &    &                           &  Zhekov et al. 
(\cite{zhekov}) \\
& & & & & \\
65 & WC9d+? & `few years' &0.1 & variable X-ray  & Oskinova \& Hamann (\cite{65}) \\
             & & & & & \\    
70  & WC9vd + B0 I &  1045$\pm$60~day  & - & photometric & Williams et al. (\cite{williams08}) \\
& & & & & \\
98a & WC8-9vd + ? & 564$\pm$8~day & - & photometric + imaging  & Monnier et al. (\cite{monnier})\\
    &             &             &   &  & Williams et al. (\cite{williams03}) \\
& & & & & \\
104 & WC9pd + B0.5 V & 241.5$\pm$0.5~day & - &  imaging & Tuthill et al. (\cite{tut08}) \\
    &                &         & & & Crowther (\cite{pac97}) \\
& & & & & \\
112 &   WC9d + ?       & 24.8$\pm$1.5~yr & - & imaging (provisional)  & Marchenko et al. (\cite{march}) \\ 
& & & & & \\
113 & WC8d+O8-9IV      & $\sim$29.707~day & - & RV   & Massey \& Niemela (\cite{mass113}) \\
& & & & & \\
118 &  WC9d + ?              & $\sim$60~day  & -    & imaging (provisional)  & Millour et al. (\cite{millour}) \\
& & & & & \\
125 & WC7ed + O9 III& $>$4~yr & - & photometric (provisional) & Williams et al. (\cite{williams94})\\
& & & & & \\
137 & WC7ed + O9  & 13.05$\pm$0.15~yr & - & RV + photometric  & Williams et al. (\cite{williams01})\\
    &   &             &          & & Lefevre et al. (\cite{lef}) \\ 
& & & & & \\
140 & WC7pd +O4-5 & 7.94$\pm$0.03~yr & 5-12 &  RV + photometric & Williams et al. 
(\cite{williams90}) \\
    & & & & & Pollock et al. (\cite{pollock}) \\
    & & & & & Zhekov \& Skinner (\cite{140}) \\
& & & & & \\
HD 36402 & WC4(+O?)+O8 I & 3.03269 & - & RV &Moffat et al. (\cite{moffatLMC}) \\
    & & & & & Williams (\cite{perry11}) \\
\hline
& & & & & \\
11 & WC8+O7.5III & 78.53$\pm$0.01~day & 1.2-1.5 & RV & Schmutz et al. (\cite{schmutz}) \\
   & & & &                   & Schild et al. (\cite{wr11}) \\
& & & & & \\
48 &WC5+O9.5/B0Iab & 19.138$\pm$0.003~day & 1.0 & RV & Hill et al. (\cite{48}) \\ 
   & (WC6+O6V) & & & & Oskinova (\cite{os05}) \\
& & & & & \\
79 &WC7+O5V & 8.8911$\pm$0.0001~day & 0.05& RV & Hill et al. (\cite{79}) \\
& & & & & Oskinova (\cite{os05}) \\
\hline
\end{tabular}
\end{center}
Column 5 summarises the origin of the orbital period, being derived from  {\bf R}adial {\bf V}elocity 
data, periodicities in the {\bf 
photometric} lightcurve, the  motion of a pinwheel nebula ({\bf imaging}) and/or a {\bf variable X-ray} flux. In several cases (WR48a, 65, 112, 118 \& 125) these are 
less robust and so are labeled as provisional. RV variations are also present in WR70 and are consistent  with both short 
($\sim$50~d) and long  ($\sim$1200~d) periods. Note the discrepant classifications given for WR48 in 
Hill et al. (\cite{48}) and Oskinva (\cite{os05}).  Naz\'{e} (\cite{93}) report an X-ray  detection of 
the WC7+O7/9 binary WR93 but no X-ray flux is given. 
\end{table*}

\subsection{W239 in an evolutionary context}

\subsubsection{Possible binary pathways}

With an orbital period of 5.05~days, W239 is one of  the most compact  WC binary system to have been unambiguously 
identified\footnote{In addition to those stars given in footnote 5, van der Hucht (\cite{cat}) lists 3 further WC stars
with shorter periods - WR 14, 50 \& 86 - in all cases solely determined from photometry with no corrobarative 
RV data. As the author cautions, the  dataset present by Shylaja (\cite{shylaja}) for WR 14 is insufficient to 
derive a period and so its identification as a binary is highly uncertain. Likewise Veen \& Wieringa 
(\cite{veen50}) cast serious doubt upon the photometric variability suggested for WR 50 by van Genderen 
et al. (\cite{vG}). Finally, Paardekooper et al. (\cite{paard}) show that the periodicity associated 
with WR86 is instead attributable to a $\beta$-Cephei companion.}, although  without dynamical mass estimates for both 
components we may not yet produce a tailored evolutionary history for it. However following the arguements of Petrovic et 
al. 
(\cite{petrovica}), Cantiello et al. (\cite{cantiello})  and De Mink et al. (\cite{casem})  we may conclude that it must 
have been
 born in a particularly compact configuration, although none of the systems simulated by these authors have
 such a short orbital period when the primary becomes a WR star (P$_{\rm orb} \geq$6~days), still less once subsequent 
mass loss has led to a  WC phase and caused the orbit to widen. Nevertheless, the WN7o+? 
P$_{\rm orb}\sim$3.51~day eclipsing binary WR B (Bonanos \cite{bonanos}) clearly demonstrates that suitable progenitor 
systems for W239 are present within Wd~1.

For evolutionary scenarios invoking  non-conservative mass transfer (e.g. Petrovic et al. \cite{petrovica}, Cantiello et al. \cite{cantiello}), 
increasing the initial orbital period of the progenitor O+O  binary results in  a progressively
more compact WR+O MS configuration. However, above a critical period and after the  transfer of 
$\sim$50\% of the envelope mass of the primary, the orbit widens and the mass transfer rate increases to the point where the 
secondary also 
expands to fill its Roche lobe (Wellstein et al. \cite{wellstein01}).  Adopting a standard common envelope prescription at this point, unpublished simulations
 by Wellstein (\cite{thesis})  found that a 16+15M$_{\odot}$, P$_{\rm orb}\sim$8~day MS binary evolves 
to a 
5+20~M$_{\odot}$, P$_{\rm orb} \sim$1.8~day configuration. Comparable  simulations with improved input physics are 
currently being  
undertaken  by Schneider et al. (in prep.), and although a higher mass progenitor  would be required 
(e.g. 30...40M$_{\odot}$ 
for a current mass of $\sim$10~M$_{\odot}$) such a common envelope pathway appears promising for the production of
 W239.

As an alternative to binary evolution driven by mass transfer we also highlight the Case M(ixing) evolutionary pathway of
 De Mink et al. (\cite{casem}). In this scenario - applicable to close, massive tidally locked binaries -   rapid  
rotation  leads to efficient rotational mixing and hence chemically homogeneous evolution. This  results in the 
stars  evolving directly to the  WR phase,  while remaining within their Roche Lobes. As with the above pathways,  
once the primary becomes a WR star, wind driven mass loss causes the initially compact orbit to widen. 
However, given that we anticipate  progenitor and current masses of $\sim$40 \& 10~M$_{\odot}$ for the WC9 component of  W239, we might expect the orbital 
period to have widened from an initial value of 1.5...2.5~days to beyond the observed $\sim$5.05~days, although confirmation of this
would require tailored modeling.

What implications do these scenarios have on the mass of the 5.05~day binary companion? Case B evolution 
leading to a common envelope phase leads to merger for low secondary masses, suggesting that it must be rather massive.
 From the unpublished thesis 
work of Wellstein  (\cite{thesis}) a time averaged mass transfer efficiency of $\sim$50\% was found for the  16+15M$_{\odot}$ binary considered. If this is 
applicable to higher mass systems such as W239 we estimate that a $\sim$40+38~M$_{\odot}$ system would result in a current 
$\sim$15+50~M$_{\odot}$ configuration, which we may exclude on observational grounds. For lower initial mass ratios one would
expect the mass transfer efficiency to be reduced (e.g. Petrovic et al. \cite{petrovica}) and so one could anticipate a 
$\sim$40+20~M$_{\odot}$ system evolving into a $\sim$15+25M$_{\odot}$ WR+MS system, broadly consistent with our current observational data. 
Weaker constraints on the mass of the binary companion are obtained for a case M
evolution, but are expected to be in the range $\sim$10--38M$_{\odot}$, although masses in the 
upper reaches of this range are again likely to be excluded on observational grounds. Nevertheless, while tailored numerical modeling
is required to confirm these estimate, it appears highly likely that the binary companion of W239 is massive enough 
 to generate  a wind collision zone and support dust production.

\subsubsection{Evolutionary pathways in Wd1}

Even if we are unable to provide a unique evolutionary scheme for W239, 
such pathways are of particular interest given that Wd~1 appears to host a number of compact WR binary systems; W239 being the 
third following WR B (WN7o+?) and W13 (WNVL+OB).
To these we might also add, albeit with less certainty,  W44 (WR L; WN9h) - which showed rapid,  dramatic line profile variability 
for which binarity is the most compelling explanation (Clark et al. \cite{clark10}) -  as well as the progenitor of the
 magnetar CXOU J164710.2-455216 (Ritchie et al. \cite{ben10}). 

However, Wd1~1 also contains a significant population of cool supergiants - 10 Yellow Hypergiants  (YHGs) \& Red Supergiants (RSGs) - and a further 3 cool B hypergiants transiting to/from such a phase (Clark et al. 
\cite{clark05}, 
\cite{clark10}). Due to  their intrinsic luminosity, these stars have such a large physical extent (e.g. $\sim$2000~R$_{\odot}$ for the RSG W26 assuming an 
intrinsic luminosity of 
log(L/L$_{\odot}$)$\sim$5.8) that  any putative companion must be in such a wide orbit that it could not influence their 
evolution.

Consequently, given the apparent coevality of  Wd~1, the twin hypergiant and short period WR binary populations provide compelling evidence for the presence of two 
distinct evolutionary pathways for $\sim$30-40M$_{\odot}$ stars.  The 
first,  experienced by single stars and long period binaries is predicted by current evolutionary theory (e.g. Meynet \& Maeder \cite{meynet})
 and results in a post-Main Sequence red loop  across the H-R diagram:\newline

{\bf Single channel:}    O6-7 V $\rightarrow$ O8-9 III $\rightarrow$ O9-B1 Ia $\rightarrow$ B5-9 Ia$^+$/YHG $\rightarrow$ 
RSG $\rightarrow$ B5-9 Ia$^+$/YHG/LBV $\rightarrow$ WN $\rightarrow$ WC\footnote{We note that the WC8 star WR K appears to be single, being amongst the faintest of the WRs within Wd1 and demonstrating  no evidence of
 RV binary reflex motion, 
an IR excess from dust emission or X-ray emission. Hence it appears   that single stars with progenitor masses $\sim$40~M$_{\odot}$ (Ritchie et al. 
\cite{ben10}) can also evolve through a WC phase prior to SNe.}  
$\rightarrow$ SN + production of black hole \newline

The second, experienced by short period binaries avoids the red loop due to binary driven mass transfer:\newline

{\bf Binary channel:} O6-7 V + OB V $\rightarrow$ WNLh +   OB III-V (e.g. W13\footnote{With a period of $\sim$9.27~days W13 will {\em not} evolve into a comparable 
configuration as WR B \& F since stellar wind losses will drive it to longer periods.}) 
$\rightarrow$ WNo + OB III-V (e.g. WR B) $\rightarrow$ WC + OB III-V (e.g. WR F)
$\rightarrow$ SN + production of neutron star  \newline

where one would predict that the wider the initial binary separation, the further through a red loop the primary travels before the onset  of mass transfer curtails it. 

The pre-SN evolutionary pathway followed also plays a significant role in the nature of the post-SN relativistic remnant. If binary driven mass loss  removes 
the H-rich mantle in - or shortly after - the main sequence, the He-rich core is exposed for a longer period of time 
than in single star evolution, allowing the enhanced WR mass loss rates to  act for longer. This  results in a lower pre-SN core mass and 
hence the likely formation of a neutron star rather than a black hole (Fryer 
et al. \cite{fryer}) as appears to be the case in Wd1 (Clark et al. \cite{clark08}; Ritchie et al. \cite{ben10}). 
In contrast, a single star will retain its 
H-rich mantle for longer, be subject to WR mass loss rates for a shorter period of time and will likely yield a Black Hole in the mass range considered here
(e.g. Wellstein \& Langer \cite{wellstein}).

Finally case M evolution, which occurs in very compact systems  (1.5...2.5~days; De Mink et al. \cite{casem}) and  proceeds without significant  binary driven mass-loss  is thought to  lead
 to rather massive pre-SNe core masses and the consequent formation of a high mass black hole.

\subsubsection{Wider implications}

It therefore appears that in addition to initial mass, metallicity and rotational velocity, binarity plays a key role in the evolution of massive stars.
If correct, what are the consequences of such an hypothesis? Firstly it may help to resolve the difficulties in reconciling the current post-MS stellar population of Wd~1 to evolutionary predictions.
Crowther et al. (\cite{pac06}) highlight this problem - specifically that the ratio of WC/WN and WN(H-rich)/WN(H-poor) WRs are discrepant - and indeed advance binarity as a possibly ammeliorative factor. 

Secondly,  the operation of two parallel  evolutionary channels would suggest that extrapolating the cluster binary fraction from the Wolf-Rayet population may be in error. If this were 
the case then one might expect the binary fraction of the unevolved O Main Sequence/Giants to be lower than that inferred for the WRs; our full  multiobject spectroscopic  dataset (Ritchie et al. 
\cite{ben09}) will allow us to test this assertion.

Beyond Wd~1, if proven, this hypothesis will have significant impact on the determination of ages and integrated masses of star forming regions. 
As an example we consider the star formation region recently detected at the base of the Scutum-Crux arm and delineated by at least four 
 red supergiant clusters (Figer et al. \cite{rsg1}; Davies et al. \cite{rsg2}; Clark et al. \cite{rsg3}; Negueruela et al. \cite{rsg4}). 
Given that extinction has prevented the identification of the Main-Sequence of each cluster, their masses have been determined via  
comparison of the number of RSGs to evolutionary predictions. If a proportion of massive stars instead avoid this phase via close binary evolution
 then integrated cluster masses and consequently star formation rates may be (substantially) underestimated (Davies et al. \cite{abund}). However, as a corollary 
one might also expect a population of  (underluminous) WR stars produced via the binary channel, which have yet to be identified. 

Indeed, the  significance  of a close binary pathway  in massive stellar evolution has already been highlighted by a number of authors.
Brinchmann et al (\cite{brinchmann}) found that binary evoluton is necessary to replicate the observational properties of Wolf-Rayet galaxies at low 
metalicities. Moreover Smith et al. (\cite{smith}) argue that the observed fraction of Type Ibc and IIb SNe
requires a large proportion of their progenitors to have evolved via  close binary evolution and  mass stripping via 
Roche Lobe overflow, while Cantiello et al. (\cite{cantiello}) demonstrate the role such a channel will play in the 
production of GRBs in lower metalicity environments. Finally, as implied by the discussion above, the prevalence of this 
pathway will also profoundly influence the production rate and properties of both intermediate- and high-mass X-ray
 binaries (e.g. Kobulnicky \& Fryer \cite{kobulnicky}), favouring the production of neutron 
star rather than black hole accretors.

\section{Conclusions}

We have analysed multi-epoch spectroscopy and photometry of the WC9d star W239, which reveal it to be a  binary with a 5.05~day period. Non-LTE modeling indicates that the parameters of the  primary are fully comparable to
 those of other Galactic WC9 stars. Inspection of the spectral energy distribution of W239 indicates the presence of hot ($\sim$1300~K) dust which, from the long 
term near-IR lightcurve  appears to be present at all epochs. These observations reveal 
substantial variability,  with a   significant near-IR `flare' between 2004-6 that is best understood as a transient 
increase in the dust production rate  resulting from the periastron passage of a massive tertiary companion in a wide, 
 eccentric orbit. 

The combination of the current stellar  census of Wd1 and the requirement that they support strong stellar  winds constrains
both secondary and tertiary components  to be  stars of spectral 
type $>$O7-8~V or  O8-9~III. Such classifications are consistent with 
the assumptions made during our non-LTE model atmosphere analysis, while out tentative identification of a Pa12 photospheric line suggests the presence
of at least one  O7-9 V-III star within the W239 system.

 W239  has one of the shortest orbital periods yet determined for {\em any} WC star.  This finding leads to several important conclusions. Firstly, it suggests 
that  dust production in WC 
stars can occur over all binary separations and is not restricted to long period systems, fully consistent with
the hypothesis that the persistent dust producing systems are all short period binaries. Future  RV observations of the complete dusty WC population within 
Wd~1 will  test this scenario.

Secondly, W239 is the third short period WR binary identified within Wd~1. The compact nature of these systems implies that they must have evolved from massive, 
short period  progenitor systems likely via non-conservative mass transfer while still on the main sequence, or shortly thereafter. In doing so they would have avoided the post-main sequence
red-loop that results in the formation of the cool YHGs and RSGs that are present within Wd~1. We therefore conclude that there is compelling evidence for the presence of two parallel 
evolutionary pathways for massive $\sim$30-40M$_{\odot}$ stars within Wd~1, depending on their binary properties (specifically their orbital separation).  In combination
with our current RV survey on OB  stars, an extention to include all remaining WRs with Wd~1 scheduled for summer 2011 will enable us to determine the relative
weighting of each channel. {\em Consequently, the  co-eval   stellar population  within Wd~1 may prove critical in determining whether binarity plays as 
significant a role in the evolution of massive stars as their  initial mass, chemical composition and 
rotational velocity are thought to.}

If this is the case, it  will have important  consequences for the nature of both GRB and SNe progenitors as well as  the subsequent production of relativistic 
objects - such as the Wd~1 magnetar - and  X-ray binaries. 
Moreover, this evolutionary bifurcation will need to be incorporated into population synthesis codes used to infer the properties of  stellar aggregates both in our own 
and external galaxies; for example, 
mass determinations for the RSG-rich clusters at the base of the Scutum-Crux arm may currently be significantly underestimated due to this omission. Finally, one might speculate that if 
binaries are predominantly 
formed via dynamical capture in stellar aggregates, rather than being primordial, then massive stellar evolution may be partially dependent on environment, with denser regions leading to 
higher capture probabilities and hence enhanced evolution via the close binary channel.

\begin{acknowledgements}

JSC acknowledges support from an RCUK fellowship. This research is partially supported by the Spanish Ministerio de 
Ciencia e Innovaci\'{o}n (MICINN) under grants AYA2008-06166-C03-03, AYA2010-21697-C05-05 and CSD2006-70 and based upon observations made at the Pico
dos Dias Observatory (LNA/MCT).  We wish to thank the referee, Peredur Williams, for his comments which greatly improved the paper and Julian Pittard 
for his valuable input into understanding the likely properties of the wind collision zone in W239. 

\end{acknowledgements}

{}

\begin{thebibliography}{}

\bibitem[1972]{allen}
Allen, D. A., Swings, J. P., Harvey, P. M., 1972, A\&A, 20, 333

\bibitem[1989]{balona}
Balona, L. A., Egan, J., Marang, F., MNRAS, 1989, 103

\bibitem[2007]{bonanos}
Bonanos, A. Z., 2007, AJ, 133, 2696

\bibitem[2005]{bonnell}
Bonnell, I. A., Bate, M. R., 2005, MNRAS, 362, 915

\bibitem[2009]{bosch}
Bosch, G., Terlevich, E., Terlevich, R., 2009, AJ, 137, 3437

\bibitem[2008]{brinchmann}
Brinchmann, J., Kunth, D., Durret, F., 2008, A\&A, 485, 657


\bibitem[2007]{cantiello}
Cantiello, M., Yoon, S.-C., Langer, N., Livio, M., 2007, A\&A, 465, L29

\bibitem[2010]{chene}
Chen\'{e}, A.-N., St-Louis, N., 2010, ApJ, 716, 929


\bibitem[2002]{clark02}
Clark, J. S., Negueruela, I., 2002, A\&A, 396, L25


\bibitem[2002]{clark02a}
Clark, J. S., Goodwin, S. P., Crowther, P. A., et al., 2002, A\&A, 392, 909

\bibitem[2005]{clark05}
Clark, J. S., Negueruela, I., Crowther, P. A., Goodwin, S. P., 2005, A\&A, 434, 949

\bibitem[2008]{clark08}
Clark, J. S., Muno, M. P., Negueruela, I., et al., 2008, A\&A, 477, 147 

\bibitem[2009]{rsg3}
Clark, J. S.,  Negueruela, I., Davies, B., et al., 2009, A\&A, 498, 109

\bibitem[2010]{clark10}
Clark, J. S., Ritchie, B. W., Negueruela, I., 2010, A\&A, 514, 87

\bibitem[2001]{cox}
Cox, A. N., 2001, Astrophysical Quantities

\bibitem[1997]{pac97}
Crowther, P. A., 1997, MNRAS, 290, L59

\bibitem[2002]{pac02}
Crowther, P. A., Dessaart, L., Hillier, D. J., Abbott, J. B., Fullerton, A. W., 2002, A\&A, 
392, 653

\bibitem[2006a]{pac06}
Crowther, P. A., Hadfield, L. J., Clark J. S., Negueruela, I., Vacca, W. D., 2006a, MNRAS, 372, 1407

\bibitem[2006b]{wc9}
Crowther, P. A., Morris, P. W., Smith, J. D., 2006b, ApJ, 636, 1033

\bibitem[2010]{ic10}
Crowther, P. A., Barnard, R., Carpano, S., et al., 2010, MNRAS, 403, L41

\bibitem[2007]{rsg2}
Davies, B., Figer, D. F., Kudritzki, R.-P., et al., 2007, ApJ, 671, 781

\bibitem[2009]{abund}
Davies, B., Origlia, L., Kudritzki, R.-P., et al., 2009, ApJ, 696, 2014

\bibitem[2000]{dem}
De Marco, O., Crowther, P. A., Schmutz, W., et al., 2000, A\&A, 358, 187
\bibitem[2009]{casem}
De Mink, S. E., Cantiello, M., Langer, N., et al., 2009, A\&A, 497, 243
\bibitem[2010]{dougherty}
Dougherty, S.M., Clark, J. S., Negueruela, I., Johnson, T., Chapman, J. M., 2010, A\&A, 511, A58

\bibitem[2006]{rsg1}
Figer, D. F., MacKenty, J. W., Robberto, M., et al., 2006, ApJ, 643, 1166

\bibitem[2002]{fryer}
Fryer, C. L., Heger, A., Langer, N., Wellstein, S., 2002, ApJ, 578, 335

\bibitem[2004]{harries}
Harries, T. J., Monnier, J. D., Symington, N. H., Kurosawa, R., 2004, MNRAS, 350, 565

\bibitem[2000]{79}
Hill,  G. M., Moffat, A. F. J., St-Louis, N., Bartzakos, P., 2000, MNRAS, 318, 402

\bibitem[2002]{48}
Hill, G. M., Moffat, A. F. J., St-Louis, N., 2002, MNRAS, 335, 1069

\bibitem[2000]{hill}
Hill, G. M., Moffat, A. F. J., St-Louis, N., Bartzakos, P., 2000, MNRAS, 318, 402

\bibitem[1998]{hillier98}
Hillier, D. J., Miller, D. L., 1998, ApJ, 496, 407

\bibitem[1999]{hillier99}
Hillier, D. J., Miller, D. L., 1999, ApJ, 519, 354


\bibitem[2001]{cat}
van der Hucht, K. A., 2001, NewAR, 45, 135

\bibitem[2001]{vdh01}
van der Hucht, K. A., Williams, P. M., Morris, P. W., 2001, ESASP, 460, 273

\bibitem[2006]{kaper}
Kaper, L., van der Meer, A., Najarro, F., 2006, A\&A, 457, 595

\bibitem[2007]{kiminki}
Kiminki, D. C., Kobulnicky, H. A., Kinemuchi, K., et al., 2007, ApJ, 664, 1102

\bibitem[2007]{kobulnicky}
Kobulnicky, H. A., Fryer, C. L., 2007, ApJ, 670, 747

\bibitem[2005]{lef}
Lef\`{e}vre, L., Marchenko, S. V., L\'{e}pine, S., et al., 2005, MNRAS, 360, 141

\bibitem[1998]{hipp}
Marchenko, S. V., Moffat, A. F. J., van der Hucht, K. A., et al., 1998, A\&A, 331, 1022

\bibitem[2002]{march}
Marchenko S. V., Moffat, A. F. J., Vacca, W. D., C\^{o}t\'{e}, S., Doyon, R., 2002,
ApJ, 565, 59

\bibitem[2007]{march07}
Marchenko, S. V., Moffat, A. F. J., 2007, ASPC, 367, 213

\bibitem[2005]{martins}
Martins, F., Schaerer, D., Hillier, D., 2005, A\&A, 436, 1049

\bibitem[2006]{plez}
Martins, F., Plez, B., 2006, A\&A, 457, 637

\bibitem[1981]{mass113}
Massey, P.,   Niemela, V. S., 1981, ApJ, 245, 195

\bibitem[2009]{mengel}
Mengel, S., Tacconi-Garman, L. E., 2009, Ap\&SS, 324, 321

\bibitem[2003]{meynet}
Meynet, G., Maeder, A., 2003, A\&A, 404, 975

\bibitem[2007]{millour}
Millour, F., Petrov, R. G.,  Chesneau, O., et al., 2007, A\&A, 464, 107


\bibitem[2009]{demink}
de Mink, S. E., Cantiello, M., Langer, N.,  et al., 2009, A\&A,  497, 243

\bibitem[1986]{moffat}
Moffat, A. F. J., Lamontagne, R., Cerruti, M., 1986, PASP, 98, 1170

\bibitem[1990]{moffatLMC}
Moffat, A. F. J., Niemela, V. S., Marraco, H. G., 1990, ApJ, 348, 232


\bibitem[1999]{monnier}
Monnier, J., Tuthill, P., Danchi, W., 1999, ApJ, 525, L97

\bibitem[1997]{morel}
Morel, T., St-Louis, N., Marchenko, V., 1997, 482, 470

\bibitem[2009]{93}
Naz\'{e}, Y., 2009, A\&A, 506, 1055

\bibitem[2010a]{rsg4}
Negueruela, I., Gonz\'{a}lez-Fern\'{a}ndez, C., Marco, A., Clark, J. S., Mart\'{i}nex-N\'{u}\~{n}ez, S., 
2010a, A\&A, 511, 840

\bibitem[2010b]{w30}
Negueruela, I., Clark, J. S., Ritchie, B. W., 2010b, A\&A, 516, A78

\bibitem[1995]{niemela}
Niemela, V. S., 1995, IAUS, 163, 223


\bibitem[2003]{os03}
Oskinova, L. M., Ignace, R., Hamann, W.-R., Pollock, A. M. T., Brown, J. C., 2003, A\&A, 402, 755

\bibitem[2005]{os05}
Oskinova, L. M., 2005, MNRAS, 361, 679

\bibitem[2008]{65}
Oskinova, L. M., Hamann, W.-R., 2008, MNRAS, 390, L78

\bibitem[1995]{gayley}
Owocki, S. P., Gayley, K. G., 1995, ApJ, 454, 1250

\bibitem[2002]{paard}
Paardekooper, S. J., Veen, P. M., van Genderen, A. M., van der Hucht, K. A., 
2002, A\&A, 384, 1012

\bibitem[1967]{pacz}
Paczynski, B., 1967, Acta Astron., 17, 355

\bibitem[2002]{pas}
Pasquani, L., Avila, G., Blecha, A., et al., 2002, The Messenger, 110, 1

\bibitem[2005a]{petrovica}
Petrovic, J., Langer, N., van der Hucht, K. A., 2005a, A\&A, 435, 1013

\bibitem[2005b]{petrovicb}
Petrovic, J., Langer, N., Yoon, S.-C., Heger, A., 2005b, A\&A, 435, 247

\bibitem[1998]{piatti}
Piatti, A. E., Bica, E., Claria, J. J.,1998, A\&AS, 127, 423

\bibitem[2009]{pittard}
Pittard, J. M., 2009, MNRAS, 396, 1743

\bibitem[1992]{pod}
Podsiadlowski, P., Joss, P. C., Hsu, J. J. L., 1992, ApJ, 391, 246

\bibitem[2005]{pollock}
Pollock, A. M. T., Corcoran, M. F., Stevens, I. R., Williams, P. M., 2005, ApJ, 629, 482

\bibitem[1976]{PU}
Prilutskii, O. F., Usov, V. V., 1976, Sov. Astr., 20, 2

\bibitem[2009]{ben09}
Ritchie, B. W., Clark, J. S., Negueruela, I., Crowther, P. A., 2009, A\&A, 507, 1585

\bibitem[2010a]{ben10}
Ritchie, B. W., Clark, J. S., Negueruela, I., Langer, N., 2010a, A\&A, 520, A48

\bibitem[2010b]{ben10b}
Ritchie, B. W., Clark, J. S., Negueruela, I., 2010b, BSRSL 80, 628 


\bibitem[2008]{sana}
Sana, H., Gosset, E., Naze, Y., Rauw, G., Linder, N., 2008, MNRAS, 386, 447

\bibitem[1992]{ml}
Schaerer, D., Maeder, A., 1992, A\&A, 263, 129

\bibitem[2004]{wr11}
Schild, H., Gudel, M., Mewe, R., et al., 2004, A\&A, 422, 177

\bibitem[1997]{schmutz}
Schmutz, W., Schweickhardt, J., Stahl, O., et al., 1997, A\&A, 328, 219


\bibitem[1990]{shylaja}
Shylaja, B. S., 1990, Ap\&SS, 164, 63

\bibitem[2008]{silverman}
Silverman, J. M., Filippenko, A. V., 2008, ApJ, 678, L17

\bibitem[2006]{skwd1}
Skinner, S. L., Simmons, A. E., Zhekov, S. A., et al., 2006, ApJ, 639, L35

\bibitem[2010]{smith}
Smith, N., Li, W., Filippenko, A. V., Chornock, R., 2010, MNRAS, in press. 


\bibitem[1999]{stevens}
Stevens, I. R., Howarth, I. D., MNRAS, 302, 549


\bibitem[1999]{tuthill99}
Tuthill, P. G., Monnier, J. D., Danchi, W. C., 1999, Nature, 398, 487

\bibitem[2008]{tut08}
Tuthill, P. G., Monnier, J. D., Lawrance, N., et al., 2008, ApJ, 675, 698


\bibitem[1991]{usov}
Usov, V. V.,  1991, MNRAS, 252, 49

\bibitem[1991]{vG}
van Genderen, A. M., Verheijen, M. A. W., van der Hucht, K. A., et al., 1991, IAUS, 143, 129

\bibitem[1998]{veen}
Veen, P. M., van der Hucht, K. A., Williams, P. M., et al., 1998, A\&A, 339, L45

\bibitem[2000]{veen50}
Veen, P. M., Wieringa, M. H., 2000, A\&A, 363, 1026

\bibitem[2000]{thesis}
Wellstein, S., 2000, PhD thesis, Potsdam University

\bibitem[1999]{wellstein}
Wellstein, S., Langer, N., 1999, A\&A, 350, 148


\bibitem[2001]{wellstein01}
Wellstein, S., Langer, N., Braun, H., 2001, A\&A, 369, 939

\bibitem[1990]{williams90}
Williams, P. M., van der Hucht, K. A., Pollock, A. M. T., et al.,
1990, MNRAS, 243, 662

\bibitem[1992]{williams92}
Williams, P. M., van der Hucht, K. A., 1992, in A. S. P. Conf. Series, Vol. 22, 269

\bibitem[1994]{williams94}
Williams, P. M., van der Hucht, K. A., Kidger, M. R., Geballe, T. R., Bouchet, P., 1994, MNRAS, 266, 247


\bibitem[2000]{williamspec}
Williams, P. M., van der Hucht, K. A., 2000, MNRAS, 314, 23


\bibitem[2001]{williams01}
Williams, P. M., Kidger, M. R., van der Hucht, K. A., et al., 2001, MNRAS, 324, 156


\bibitem[2003]{williams03}
Williams, P. M., van der Hucht, K. A., Morris, P. W., Marang, F., 2003, IAUS, 212, 115

\bibitem[2008]{williams08}
Williams, P. M., 2008, RevMexAA, 33, 71

\bibitem[2009]{williams09}
Williams, P. M., Marchenko, S. V., Marston, A. P., et al., 2009, MNRAS, 395, 1749

\bibitem[2011]{perry11}
Williams, P. M., 2011, BSRSL, 80, 195


\bibitem[2000]{140}
Zhekov, S. A., Skinner, S. L., 2000, ApJ, 538, 808

\bibitem[2010]{zhekov}
Zhekov, S A., Gagn\'{e}, M., Skinner, S. L., ApJL, 2011, 727, L17

\bibitem[2007]{zin}
Zinnecker, H., Yorke, H. W., 2007, ARA\&A, 45, 481

\end{thebibliography}
\end{document}